\documentclass[twocolumn]{aastex631} 

\usepackage{makecell}
\usepackage{booktabs}
\usepackage{hyperref}
\usepackage{nameref}
\usepackage{lineno}
\usepackage{url}   
\usepackage{appendix}  

\hypersetup{linkcolor=red,citecolor=cyan,filecolor=blue,urlcolor=magenta}
\usepackage{subfigure}
\usepackage{color}
\shorttitle{GeV Flaring Activity of the Young Radio Galaxy 4C 76.03}
\shortauthors{Jiang et al.}

\begin{document}

\title{The First GeV Gamma-Ray Flares from the CSO-like Source 4C 76.03}

\correspondingauthor{Da-Ming Wei}
\email{ dmwei@pmo.ac.cn}

\author[0009-0004-3221-2603]{Xiong Jiang}
\affiliation{Key Laboratory of Dark Matter and Space Astronomy, Purple Mountain Observatory,
Chinese Academy of Sciences, Nanjing 210023, People’s Republic of China}
\affiliation{School of Astronomy and Space Science, University of Science and Technology of China, Hefei, Anhui 230026, People’s Republic of China}

\author[0009-0005-1848-0553]{Hai Lei}
\affiliation{School of Physics and Materials Science, Guangzhou University, Guangzhou 510006, People’s Republic of China}

\author{Hao-Yi Huang}
\affiliation{Department of Physics and Astronomy, College of Physics, Guizhou University, Guiyang 550025, China}

\author{Wei Zhang}
\affiliation{Department of Physics and Astronomy, College of Physics, Guizhou University, Guiyang 550025, China}

\author[0009-0002-4513-4486]{Yang-Ji Li}
\affiliation{Yunnan Observatories, Chinese Academy of Sciences, Kunming 650216, People’s Republic of China}
\affiliation{University of Chinese Academy of Sciences,  Beijing 100049, People’s Republic of China}

\author[0000-0002-9758-5476]{Da-Ming Wei}
\affiliation{Key Laboratory of Dark Matter and Space Astronomy, Purple Mountain Observatory,
Chinese Academy of Sciences, Nanjing 210023, People’s Republic of China}
\affiliation{School of Astronomy and Space Science, University of Science and Technology of China, Hefei, Anhui 230026, People’s Republic of China}


\begin{abstract}
We report the first detection of GeV $\gamma$-ray flaring activity from the compact symmetric object (CSO)–like source 4C 76.03, based on 17 years of \emph{Fermi}-LAT observations. Its long-term, time-averaged $\gamma$-ray properties are consistent with the 4FGL-DR4 catalog. However, a time-resolved analysis with 100-day binning reveals two prominent flares occurring on timescales of approximately 30 days and 20 days, separated by $\sim2.5$~yr, with nearly identical fluxes, test statistic (TS) values, and photon indices. The short-timescale variability indicates localized and transient energy dissipation in the nuclear region, likely associated with newly injected jet components. Although the $\gamma$-ray emission does not directly trace the long-term jet power responsible for building the observed radio structure, it demonstrates that the central engine remains active. In the context of CSO evolution, 4C 76.03 may represent a rare transitional case, where repeated energy injections allow the source to exceed the canonical $\sim$500~pc scale of most CSOs, providing key insight into the early stages of radio jet evolution.
\end{abstract}


\section{Introduction} \label{sec:intro}

The detection of $\gamma$-ray emission in galaxies indicates the presence of extreme physical conditions. Among extragalactic sources, active galactic nuclei (AGNs) are prominent $\gamma$-ray emitters, powered by accretion onto central supermassive black holes (SMBHs) \citep{1993ARA&A..31..473A}. Approximately 10\% of AGNs launch powerful relativistic jets, classified as radio-loud AGNs \citep{1995PASP..107..803U, 2019ARA&A..57..467B}. When these jets are closely aligned with our line of sight, the sources are observed as blazars, whose high-energy emission is strongly amplified by Doppler boosting, dominating the extragalactic $\gamma$-ray sky \citep{2015A&ARv..24....2M, 2023arXiv230712546B}. The latest incremental release of the Fourth Fermi-LAT AGN Catalog (4LAC-DR3; \citealt{2022ApJS..263...24A}) lists 3,405 AGNs at Galactic latitudes $|b|>10^\circ$, with $\sim$98\% classified as blazars and only $\sim$2\% as radio galaxies. The latter are generally considered the misaligned counterparts of blazars, with larger jet inclination angles and consequently weaker Doppler boosting \citep{1974MNRAS.167P..31F, 1984ARA&A..22..319B}. Their misaligned jets provide a unique opportunity to probe extreme non-beamed high-energy processes in AGNs, typically obscured by the strongly beamed emission in blazars \citep{2018A&A...614A...6S,2021MNRAS.507.4564P}.

Radio AGNs exhibit a broad range of projected linear sizes, from parsec scales up to several Mpc. The largest, classical radio galaxies are typically classified as Fanaroff–Riley Type I (FR I) or Type II (FR II) \citep{1974MNRAS.167P..31F}. A population of compact radio sources often shows a spectral turnover due to synchrotron self-absorption or free-free absorption \citep{1997AJ....113..148O}. Based on their spectra, these small yet powerful sources are classified as GHz-peaked spectrum (GPS) or compact steep spectrum (CSS) sources \citep{1991ApJ...380...66O,1997ApJ...485..112B,2021A&ARv..29....3O}. Morphologically, GPS and CSS sources resemble miniature versions of double-lobed FR II galaxies, with symmetric lobes on either side of a faint radio core. According to the standard evolutionary scenario, these compact sources are expected to grow into large-scale radio galaxies \citep{1996ApJ...460..634R,2000MNRAS.319..445S,2014MNRAS.438..463O}. Based on projected linear size (LS), they are further classified as compact symmetric objects (CSOs) if LS $\lesssim 1$~kpc, medium symmetric objects (MSOs) if LS $\sim 1$–20~kpc, and large symmetric objects (LSOs) if LS $> 20$~kpc \citep{2013MNRAS.433..147D}.

Being the smallest and most compact radio sources, CSOs are crucial for understanding the evolution of relativistic jets \citep{2021AN....342.1185R}. They likely represent an early evolutionary stage of radio galaxies, with kinematic ages $\lesssim$ a few thousand years. Their compactness may result from youth, a dense galactic environment that hinders jet propagation, and/or episodic nuclear jet activity \citep{2021A&ARv..29....3O}. To obtain a robust sample of CSOs, \citet{2024ApJ...961..240K} recently compiled a catalog of 79 bona fide CSOs based on literature review and multifrequency radio observations. These sources satisfy: (i) projected radio structure length $< 1$~kpc, (ii) radio emission detected on both sides of the active center, (iii) fractional variability $\le 20\%~\mathrm{yr}^{-1}$, and (iv) no superluminal motion exceeding $v_{\mathrm{app}} = 2.5c$. CSOs can be divided into two fundamentally distinct types: a low-luminosity, ``edge-dimmed'' class (CSO-1) and a high-luminosity, ``edge-brightened'' class (CSO-2) \citep{2016MNRAS.459..820T,2024ApJ...961..240K}. Other studies have attempted to confirm more morphologically complex CSO candidates \citep{2025ApJ...987...26S,2025A&A...704A..93A}.

Similar to classical radio galaxies, a small number of CSOs have been detected with significant $\gamma$-ray emission, including NGC~6328 (associated with 4FGL~J1724.2$-$6501; \citealt{2016ApJ...821L..31M}), TXS~0128+554 (associated with 4FGL~J0131.2+5547; \citealt{2020ApJ...899..141L}), NGC~3894 (counterpart of 4FGL~J1149.0+5924; \citealt{2020A&A...635A.185P}), 4C~+39.23B \citep{2022ApJ...927..221G}, DA~362 (associated with 4FGL~J1416.0+3443; \citealt{2025ApJ...979...97S}) and JVAS J1311+1658 \citep{2026ApJ...999..182J}. Their compact radio lobes are expected to contain abundant highly relativistic particles, with $\gamma$-ray emission potentially produced via interactions between these electrons and low-energy optical--UV photons from the accretion disk \citep{2008ApJ...680..911S}.

In this paper, we report on the flaring behavior of $\gamma$-ray emission from the young radio galaxy 4C~76.03 using \emph{Fermi}-LAT observations. Evidence for possible $\gamma$-ray emission from this source was previously reported by \citep{2021MNRAS.507.4564P}, who obtained TS $\sim$ 12 (corresponding to $\gtrsim 3\sigma$ significance) based on more than 11 years of \emph{Fermi}-LAT data.  In the latest 14-year \emph{Fermi}-LAT catalog \citep{2023arXiv230712546B, 2022ApJS..260...53A}, this source is potentially associated with a low-significance ($\sim 4.3\sigma$) source, 4FGL~J0410.6+7656. 4C~76.03 was among the first discovered CSO-2 sources \citep{1996ApJ...460..612R}, and our study of its $\gamma$-ray behavior supports the recent scenario proposed by \citet{2024ApJ...961..242R}: contrary to the traditional evolutionary model, most CSO-2s are short-lived sources and do not complete a full life cycle from birth to large-scale radio galaxies. Only a small fraction can grow into large-scale radio galaxies, with 4C~76.03 representing a rare example. We adopt a $\Lambda$CDM cosmology with $\Omega_M = 0.32$, $\Omega_\Lambda = 0.68$, and a Hubble constant $H_0 = 67$ km s$^{-1}$ Mpc$^{-1}$ \citep{2014A&A...571A..16P}.

\section{Data Analysis and Results} \label{sec:2}
\subsection{Fermi-LAT Data and Analysis} \label{subsec:1}

The \emph{Fermi} Large Area Telescope (LAT) is a pair-conversion $\gamma$-ray detector operating over an energy range from $\sim$20~MeV to beyond 300~GeV. It consists of a silicon-strip tracker interleaved with tungsten converter foils, a CsI(Tl) calorimeter for energy measurements, and an anticoincidence detector for charged-particle background rejection \citep{2009ApJ...697.1071A}.

In this work, we analyze $\sim$17~yr of \emph{Fermi}-LAT Pass~8 data collected between 2008 August 5 and 2025 September 14 (MJD 54683--60932) in the energy range 0.1--300~GeV. SOURCE-class events were selected ({\tt evclass}=128, {\tt evtype}=3; \citealt{2018arXiv181011394B}). To reduce contamination from Earth-limb $\gamma$-rays, events with zenith angles larger than $90^{\circ}$ were excluded. Time intervals during which the LAT was not operating in nominal science mode were removed by applying the recommended good-time-interval selections ({\tt DATA\_QUAL==1 \&\& LAT\_CONFIG==1}). The data reduction and analysis were performed using the \texttt{Fermitools}\footnote{\url{https://fermi.gsfc.nasa.gov/ssc/data/analysis/documentation/}} (version~2.0.8) together with the corresponding \texttt{Fermitools-data} package (v0.18).

A circular region of interest (ROI) with a radius of $10^{\circ}$ was defined, centered on the radio position of the source 4C~76.03, which is spatially associated with the cataloged \emph{Fermi}-LAT source 4FGL~J0410.6+7656 in the 4FGL-DR4 catalog \citep{2023arXiv230712546B}. 
The background model includes all 4FGL-DR4 sources within a $15^{\circ}$ radius of 4FGL~J0410.6+7656, as well as the Galactic diffuse emission model ({\tt gll\_iem\_v07}) and the isotropic diffuse component ({\tt iso\_P8R3\_SOURCE\_V3\_v01})\footnote{Available at \url{https://fermi.gsfc.nasa.gov/ssc/data/access/lat/BackgroundModels.html}}. 

The normalizations of the diffuse components were allowed to vary during the likelihood fitting. 
For sources outside $10^{\circ}$ of the ROI center, all  parameters were fixed to their 4FGL-DR4 catalog values\footnote{\url{https://fermi.gsfc.nasa.gov/ssc/data/access/lat/14yr_catalog/}}.
For sources within $10^{\circ}$ of the ROI center, a two-step fitting strategy was adopted for their spectral parameters. 
In the first step, the spectral indices of the background sources were fixed to their catalog values, and only the normalizations were allowed to vary.
The detection significance was quantified using the TS value, defined as
\begin{equation}
\mathrm{TS} = -2 \ln \left( \frac{L_0}{L} \right),
\end{equation}
where $L_0$ and $L$ are the maximum likelihoods of the model without and with the source included, respectively \citep{1996ApJ...461..396M}.
In the subsequent fit, the spectral indices of sources with TS $>25$ within $3^{\circ}$ of the ROI center and TS $>50$ beyond $3^{\circ}$ were freed, 
while the remaining sources were kept fixed to their catalog values.

The $\gamma$-ray emission from 4FGL J0410.6+7656 was modeled using a power-law spectrum of the form
\begin{equation}
\frac{dN}{dE} = N_0 \left( \frac{E}{E_0} \right)^{-\Gamma}.
\end{equation}
We performed a binned likelihood analysis to study the long-term, $\sim$17-year time-averaged $\gamma$-ray properties of 4FGL~J0410.6+7656, and a time-resolved, unbinned likelihood analysis using 100-day bins to capture its short-timescale variability. The unbinned approach allows us to maximize the use of all available photon information, particularly in bins with low photon counts.
For time bins in which the source was significantly detected (TS $>25$), fluxes and spectral parameters were derived using an unbinned maximum-likelihood analysis. 
Otherwise, for intervals with low detection significance (TS $<10$), 95\% confidence-level (C.L.) upper limits were derived using the \texttt{UpperLimits} module in \texttt{pyLikelihood}.
We generated a residual TS map for the $\sim$17-year time-averaged analysis to identify sources not included in the 4FGL-DR4 catalog, and reran the likelihood fit with an updated background model whenever such sources were found.
In the time-resolved analysis, the parameters of background sources with TS $<$ 10 were fixed to the values from the full 17-year analysis to ensure a stable likelihood fit.

\subsection{Fermi-LAT Result} \label{subsec:2}

In 4FGL-DR4, based on 14~yr of \emph{Fermi}-LAT observations, the source is characterized by $\mathrm{TS}\sim25$, an integrated photon flux of $\sim(4.8 \pm 1.3)\times10^{-9}\ \mathrm{ph\ cm^{-2}\ s^{-1}}$, and a photon index of $\Gamma = 2.77 \pm 0.21$. 
Using the full $\sim$17 yr \emph{Fermi}-LAT data set, we obtained a test statistic of $\mathrm{TS}=26.4$, together with an integrated photon flux of $(4.6 \pm 1.2)\times10^{-9}\ \mathrm{ph\ cm^{-2}\ s^{-1}}$. 
The $\gamma$-ray spectrum is well described by a power-law model with a photon index of $\Gamma = 2.80 \pm 0.26$, consistent within uncertainties with the catalog values.

\begin{figure*}
    \centering
    \includegraphics[scale=0.8]{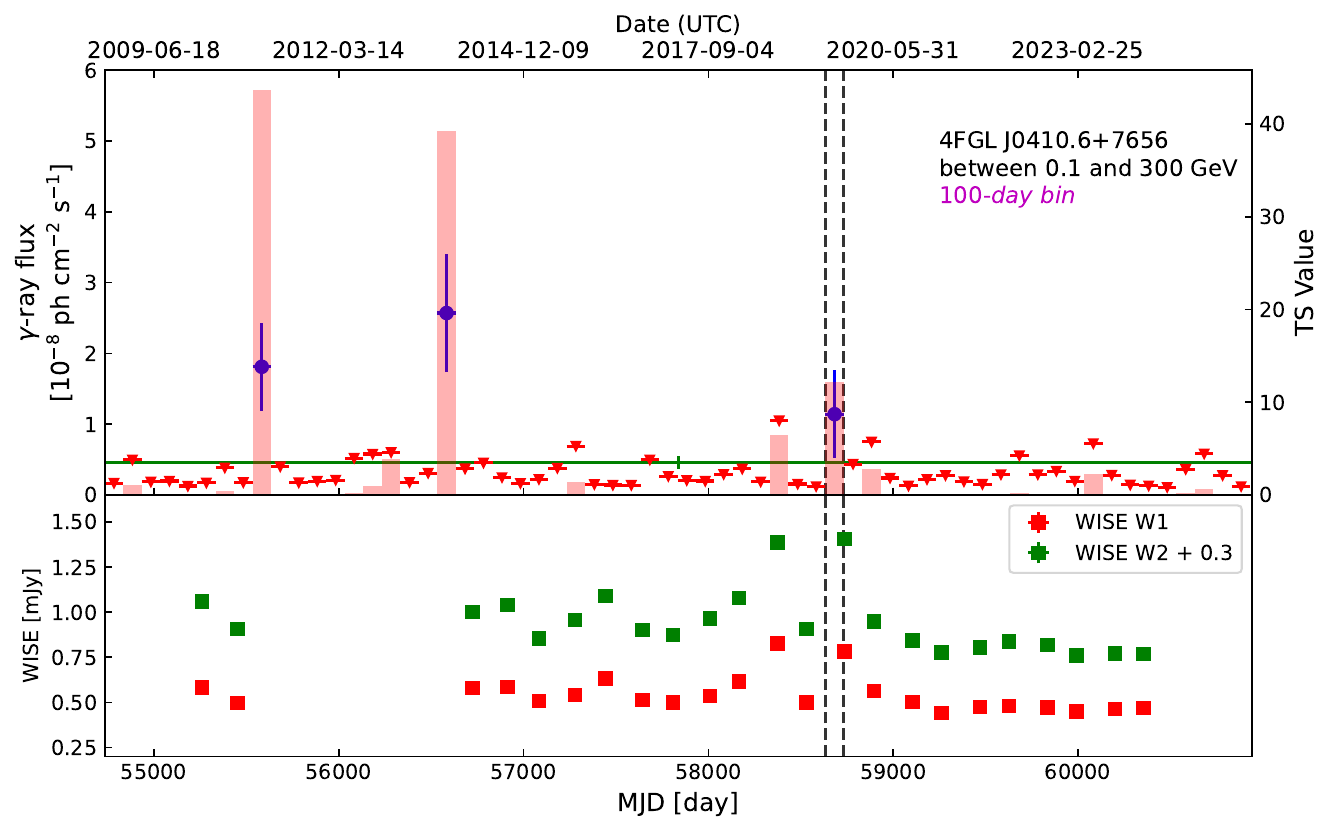}

    \begin{minipage}{0.48\textwidth}
        \centering
        \includegraphics[width=\textwidth]{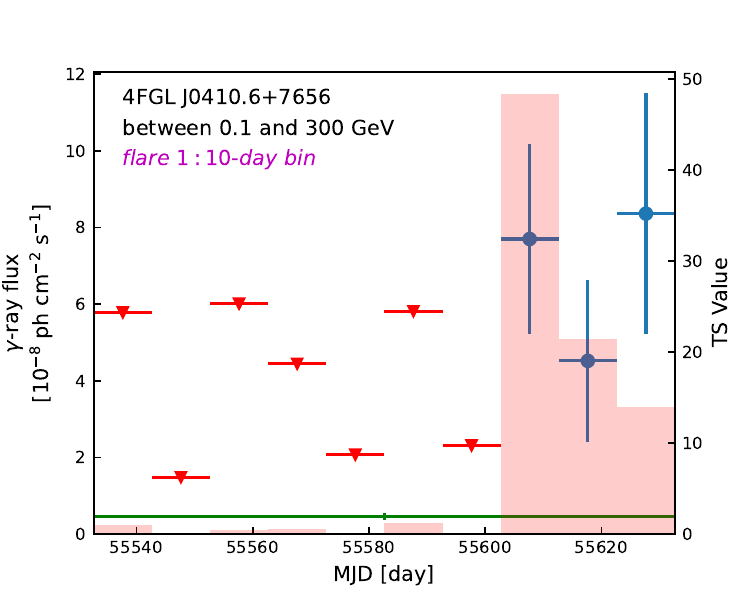}
    \end{minipage}
    \hfill
    \begin{minipage}{0.48\textwidth}
        \centering
        \includegraphics[width=\textwidth]{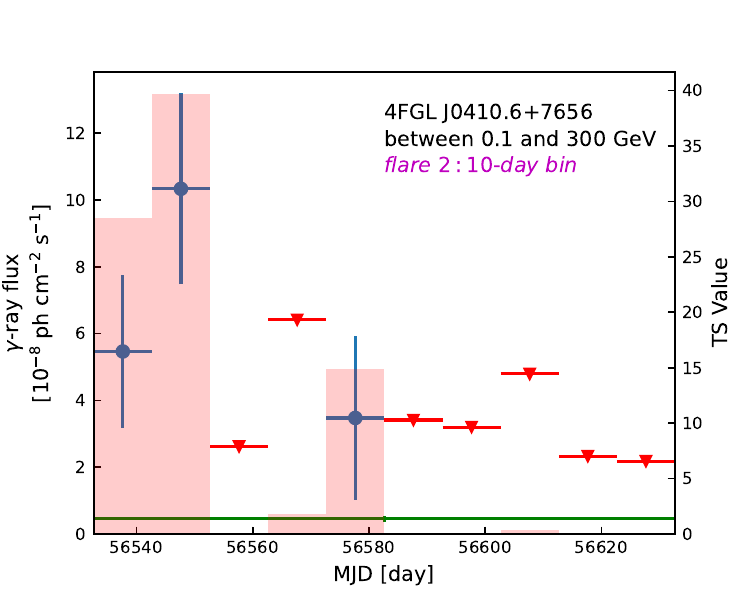}
    \end{minipage}

    \caption{
    \textbf{Upper panel:} 100-day binned $\gamma$-ray light curve of the target source, covering MJD 54732.6 to MJD 60932.6 (approximately 17 years of \emph{Fermi}-LAT observations). Blue circles represent the $\gamma$-ray flux measurements, red triangles indicate flux upper limits (TS $<$ 10), and the semi-transparent red bars show the corresponding TS values. The green horizontal line indicates the 17-year long-term average flux. The black dashed lines highlight a possible minor flaring episode during MJD 58632.6--58732.6. The lower portion shows the WISE W1 and W2 band light curves of the source.  
    \textbf{Bottom panels:} {\it Left:} Zoomed-in 10-day binned view of the 100-day binned $\gamma$-ray light curve for the first flare (MJD 55532.6--55632.6). {\it Right:} Zoomed-in 10-day binned view of the 100-day binned $\gamma$-ray light curve for the second flare (MJD 56532.6--56632.6).
}

    \label{lightcurve}
\end{figure*}

We then constructed a $\gamma$-ray light curve with a bin size of 100~days, as shown in Figure \ref{lightcurve}. Two prominent $\gamma$-ray flares appear in the light curve. Outside these two intervals, the source remains in a low-flux state for
the majority of the time and is not significantly detected by the
\emph{Fermi}-LAT.
 This suggests that the $\gamma$-ray detection of 4FGL~J0410.6+7656 is largely driven by these two 100-day flaring episodes.
To further investigate the variability behavior of this source, we constructed light curves with a finer binning of 10~days for each of the two 100-day flaring intervals. The resulting light curves are also shown in Figure \ref{lightcurve}. The first flare is primarily contributed by enhanced emission during the latter $\sim$30~days of the 100-day interval (MJD~55602.6--55632.6), whereas the second flare is mainly associated with increased emission during the first $\sim$20~days of the corresponding 100-day interval (MJD~56532.6--56552.6).

\begin{figure*}
    \centering
    \includegraphics[scale=0.45]{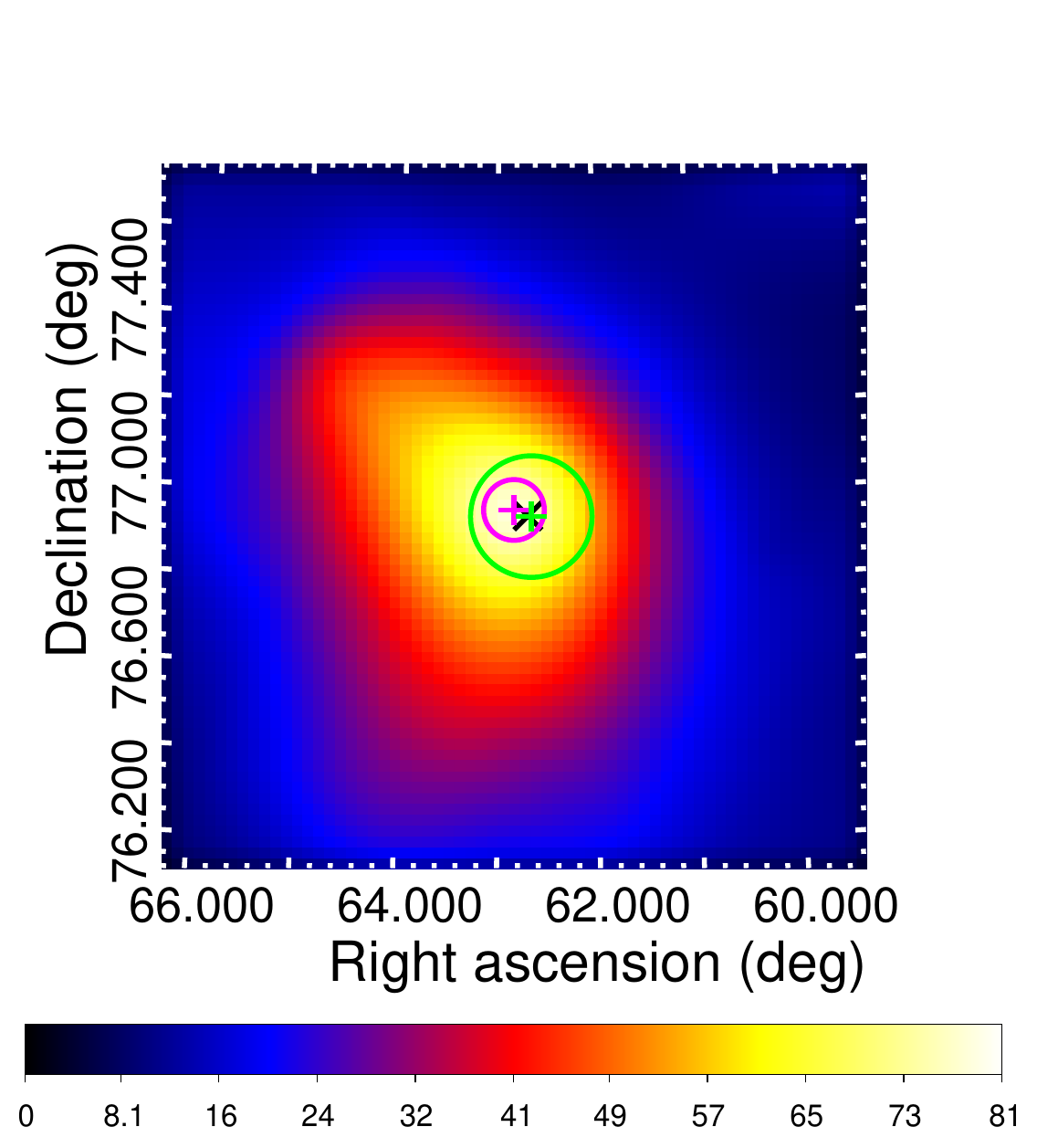}
    \includegraphics[scale=0.45]{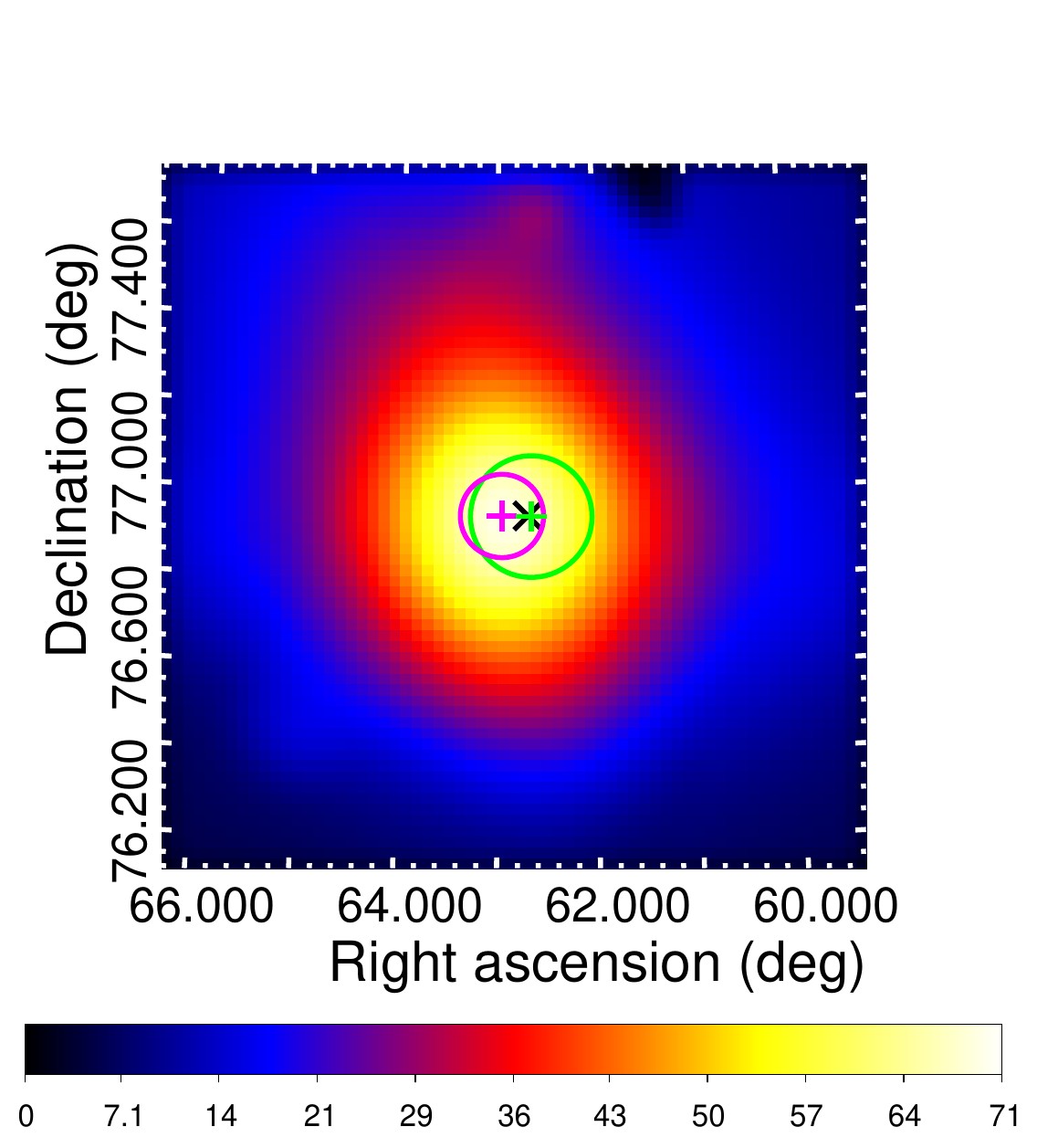}
    \includegraphics[scale=0.5]{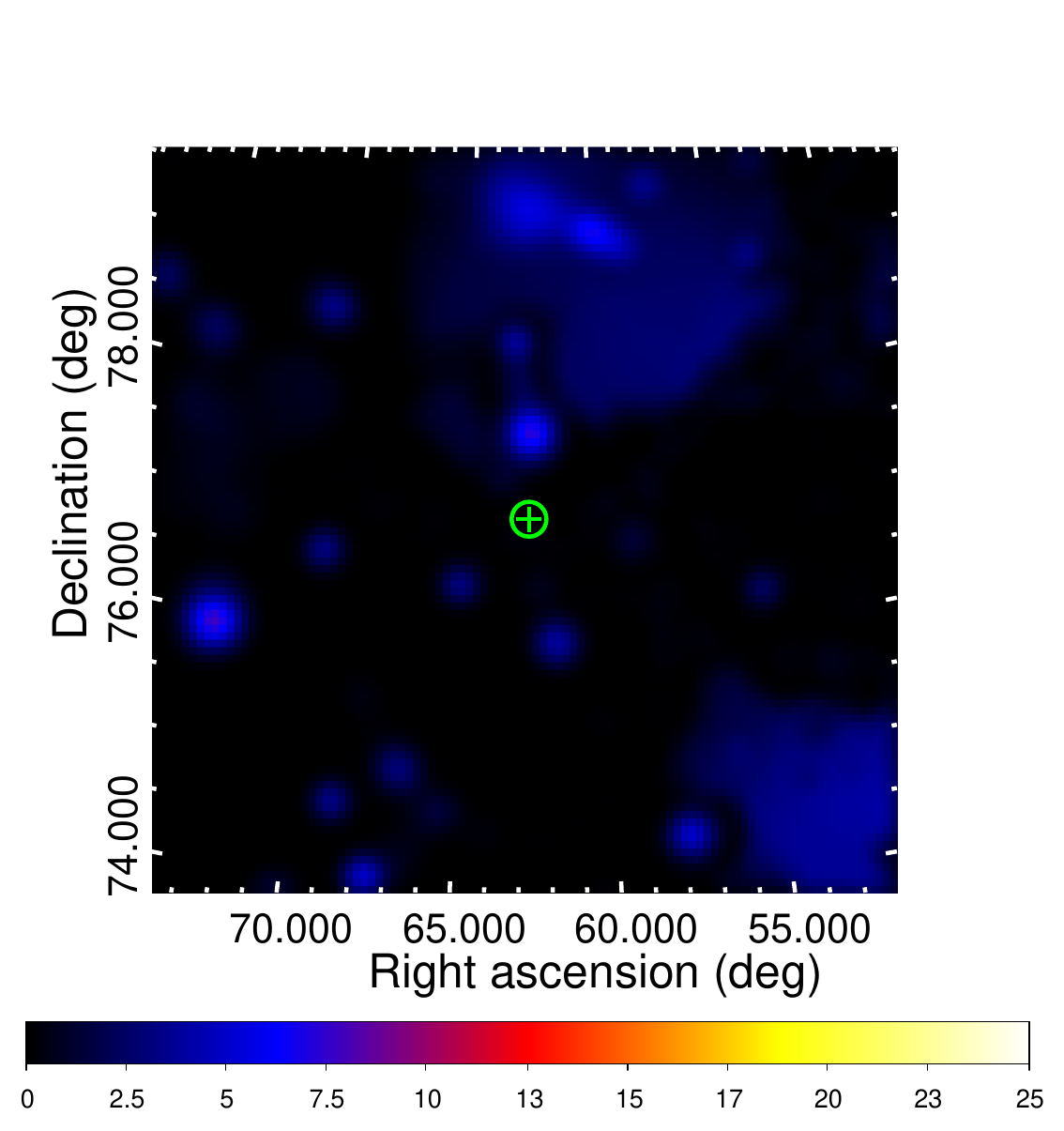}
    \caption{
    \textbf{Upper panels:} Smoothed $\gamma$-ray TS maps in the 100~MeV--300~GeV energy range, with the target source excluded from the background model. 
    {\it Left panel:} TS map derived from 30 days of \emph{Fermi}-LAT data corresponding to the first flare (MJD~55602.6--55632.6). 
    The magenta cross and circle indicate the best-fit position of the target $\gamma$-ray source and its 95\% C.L. localization error circle, respectively. 
    The black X marks the radio position of 4C~76.03, while the green cross and circle indicate the best-fit position of the target $\gamma$-ray source and its 95\% C.L. localization error circle obtained from the $\sim$17-yr data set. 
    {\it Right panel:} TS map derived from 20 days of \emph{Fermi}-LAT data corresponding to the second flare (MJD~56532.6--56552.6). 
    The symbols have the same meanings as in the left panel. 
    \textbf{Bottom panel:} Residual TS map based on $\sim$17-yr of \emph{Fermi}-LAT data, with the target source included in the model. 
    The symbols have the same meanings as in the upper panels.
}

    \label{tsmap}
\end{figure*}

\begin{figure*}
    \centering
    \includegraphics[scale=0.8]{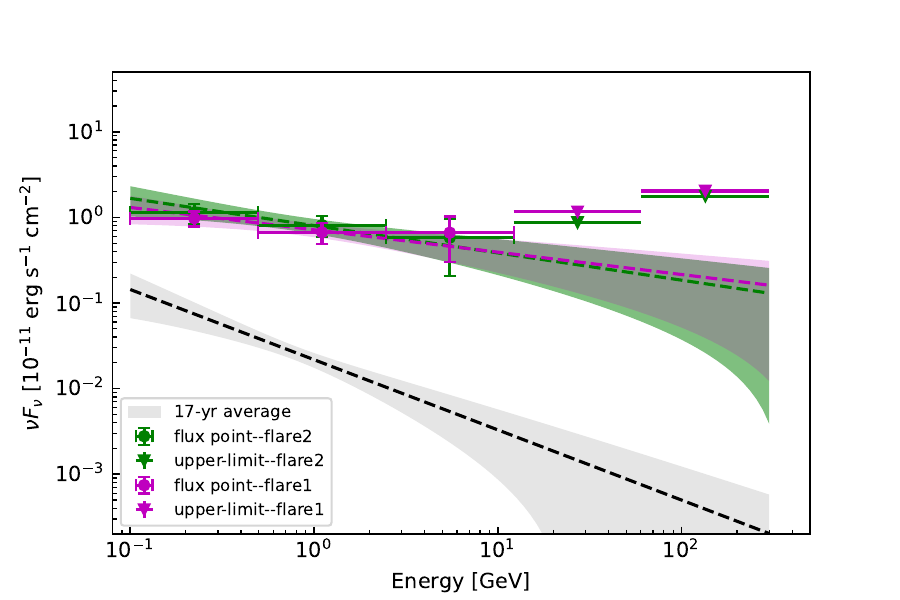}
    \caption{
The $\gamma$-ray SED of the target source in the 100~MeV--300~GeV energy range.
Circular data points represent the energy flux, while triangles indicate upper limits on the energy flux (TS $< 10$).
The dashed line shows the best-fit model, and the shaded region denotes the $1\sigma$ uncertainty range.
The purple and green colors correspond to two different flaring periods, while the gray color represents the results from approximately 17 years of data.
}

    \label{gamma_SED}
\end{figure*}

For the 30-day interval corresponding to the first flaring episode, we obtained a test statistic of 
$\mathrm{TS}=81.2$, and the corresponding TS map is shown in Figure~\ref{tsmap}. The same figure also includes the residual TS map derived from the full $\sim$17-yr dataset, which shows no evidence for previously uncataloged sources around 4FGL J0410.6+7656.
In the 0.1--300~GeV energy range, the integrated photon flux is 
$(6.5 \pm 1.5)\times10^{-8}\ \mathrm{ph\ cm^{-2}\ s^{-1}}$, 
approximately 14 times higher than the average flux derived from the full 17~yr data set. 
The photon index during this interval is $\Gamma = 2.26 \pm 0.15$, which is harder than the value obtained from the 17~yr averaged analysis. 
In addition, we estimated the corresponding energy flux to be $(4.4 \pm 1.0)\times10^{-11}\ \mathrm{erg\ cm^{-2}\ s^{-1}}$, which is approximately 25 times higher than the 17-yr averaged energy flux, estimated to be $\sim1.7\times10^{-12}\ \mathrm{erg\ cm^{-2}\ s^{-1}}$, indicating a pronounced high-energy enhancement. The resulting spectral energy distribution (SED) is shown in Figure~\ref{gamma_SED}.

Given the relatively low detection significance of 4FGL~J0410.6+7656 in 4FGL-DR4 (at the level of $\sim$4.3$\sigma$), 
we performed an additional source localization analysis using the \texttt{gtfindsrc} tool. 
This analysis yielded a best-fit position at 
R.A. = $62.83^\circ$ and Dec. = $76.96^\circ$, 
with a 95\% C.L. positional uncertainty of $0.07^\circ$, 
in agreement, within uncertainties, with the position reported in 4FGL-DR4.
Within the 95\% confidence localization region, 4C~76.03 is the only radio source listed in the Radio Fundamental Catalog \citep{2025ApJS..276...38P}, 
with an angular separation of $0.035^\circ$ from the best-fit $\gamma$-ray position. 
A Bayesian association analysis performed with \texttt{gtsrcid} yields an association probability of 99.1\%, 
suggesting that 4C~76.03 is the most plausible radio counterpart of 4FGL~J0410.6+7656. 
This identification is further supported by the 4FGL-DR4 catalog, which reports an association probability of 99.7\% 
between 4C~76.03 and 4FGL~J0410.6+7656.

A similar analysis was carried out for the 20-day interval corresponding to the second flaring episode. 
During this period, we obtained a test statistic of $\mathrm{TS}=71.3$. 
In the 0.1--300~GeV energy range, the integrated photon flux is 
$(8.0 \pm 1.9)\times10^{-8}\ \mathrm{ph\ cm^{-2}\ s^{-1}}$, 
corresponding to an energy flux of 
$(4.8 \pm 1.1)\times10^{-11}\ \mathrm{erg\ cm^{-2}\ s^{-1}}$. 
The photon index during this interval is $\Gamma = 2.32 \pm 0.16$. 
An additional localization analysis using the \texttt{gtfindsrc} tool yielded a best-fit position at 
R.A. = $62.95^\circ$ and Dec. = $76.95^\circ$, 
with an approximate 95\% C.L. positional uncertainty of $0.10^\circ$.
Overall, the results obtained for this 20-day interval are broadly consistent with those derived for the 30-day interval of the first flaring episode, with comparable TS values, $\gamma$-ray fluxes, spectral indices, and source localization results.

In addition to the two major flaring episodes on ~20-day timescales discussed above, we note a possible weaker activity episode in the $\gamma$-ray light curve constructed with 100-day time bins. In particular, bin 40 (MJD 58632.6–58732.6) shows a mild enhancement in the $\gamma$-ray flux. Variability on timescales shorter than 100 days cannot be constrained by the current analysis due to limited photon statistics.
During this interval, 4C~76.03 is detected with a test statistic of 
$\mathrm{TS} = 12.1$. 
The integrated photon flux in the 0.1--300~GeV energy range is 
$(1.2 \pm 0.62)\times10^{-8}\ \mathrm{ph\ cm^{-2}\ s^{-1}}$, 
and the photon index is $\Gamma = 2.38 \pm 0.27$. 
Within uncertainties, the derived spectral index is consistent with those obtained 
during the two earlier major flaring episodes. 
However, given the relatively low detection significance, 
this possible long-duration activity should be interpreted with caution.

\subsection{Other Observations} \label{subsec:3}
4C~76.03 (0404+768) is optically identified with a galaxy at a redshift of
$z = 0.5985$ \citep{1991ApJ...380...66O}, and its radio spectrum exhibits a low-frequency turnover at
$\nu_{\rm p} \simeq 0.55$~GHz \citep{2014MNRAS.438..463O}.
The radio emission is dominated by non-core components, with a core radio luminosity of
$\log L_{\rm core} = 26.43\ {\rm W\ Hz^{-1}}$ compared to a total radio luminosity of
$\log L_{\rm tot} = 28.08\ {\rm W\ Hz^{-1}}$, while the extended emission shows a steep spectrum
($\alpha \simeq 0.5$) and the core exhibits an inverted spectrum ($\alpha \sim -0.5$) \citep{2013MNRAS.433..147D}. The source shows strong depolarization between 15 and 8.0~GHz, with a lower fractional polarization at 8.4~GHz likely affected by uncertainties due to the intrinsically low polarization level \citep{2013A&A...555A...4M}.

4C~76.03 appears to be faint in the optical and ultraviolet bands.
According to the most recent radio fundamental catalog compiled by \citep{2025ApJS..276...38P}, the source is listed in the Pan-STARRS catalog \citep{2016arXiv161205560C}, with reported magnitudes of $g = 23.1$, $r = 21.6$, $i = 20.8$, and $z = 20.3$~mag.
We attempted to extract an optical light curve from the Zwicky Transient Facility (ZTF; \citet{2019PASP..131a8003M}) survey; however, the available data are of insufficient quality for a reliable variability analysis.

Despite its extreme faintness at optical wavelengths, 4C~76.03 is clearly detected in the infrared bands observed by the Wide-field Infrared Survey Explorer (WISE; \citet{2010AJ....140.1868W}) and the Two Micron All-Sky Survey(2MASS; \citet{2006AJ....131.1163S}).
In the AllWISE catalog \citep{2013wise.rept....1C}, the source has magnitudes of $W1 = 14.383$, $W2 = 13.489$, and $W3 = 10.589$~mag.
We extracted the WISE $W1$ and $W2$ light curves of the source, as shown in Figure~\ref{lightcurve}.
Unfortunately, no WISE observations are available during the two major $\gamma$-ray flaring episodes.
During a later interval coincident with a possible low-level $\gamma$-ray activity phase, the infrared emission shows a modest enhancement, with the $W1$ and $W2$ flux densities increasing by approximately 56\% and 82\%, respectively, relative to the preceding epoch.

\section{Discussion and conclusion} \label{sec:4}

Recent studies \citep{2024ApJ...961..240K,2024ApJ...961..241K} have identified 17 spectroscopically confirmed CSO-2 objects from three statistically complete radio samples: the Pearson--Readhead sample \citep{1981ApJ...248...61P, 1988ApJ...328..114P}, the first Caltech--Jodrell Bank sample \citep{1995ApJS...98....1P, 1995ApJS...99..297X}, and the Peacock \& Wall sample \citep{1981MNRAS.194..331P, 1985MNRAS.216..173W}. Analysis of their projected linear sizes reveals a pronounced upper cutoff at $\sim 500$--$600$~pc ($p = 1.7 \times 10^{-4}$, $\sim 3.6\sigma$), indicating a physical limit on the growth of most CSO-2 jets, corroborated by independent findings \citep{2024ApJ...977..195D}. Although 4C~76.03 was initially classified as a CSO-2 \citep{1996ApJ...460..612R}, its linear size exceeds this cutoff by roughly 20\%, leading to its exclusion from the most recent CSO-2 sample \citep{2024ApJ...961..241K,2024ApJ...961..242R}. Nevertheless, it retains several characteristic features of CSO-2 sources, and we classify it as a CSO-2-like object to emphasize its similarity to typical CSO-2s in morphology and evolutionary properties.

The existence of an upper size limit for CSO-2s suggests physical constraints on their jet formation. According to \citet{2024ApJ...961..242R}, most CSO-2s are powered not by long-lived stochastic jets, but by short-lived, high-luminosity outflows triggered by tidal disruption events (TDEs), in which a star partially disrupted by a supermassive black hole launches jets lasting $\sim 10^2$--$10^3$~yr. Higher-energy CSO-2s may require more extreme TDEs, involving rapidly spinning black holes, relatively stable magnetic fields, and disk--wind cycling to sustain jet activity and enhance energy output. This scenario naturally explains the compactness, high symmetry, and general failure of most CSO-2s to evolve into medium- or large-scale radio galaxies. The rate of TDEs is approximately 
$\sim 10^3\ \mathrm{Gpc^{-3}\ yr^{-1}}$ \citep{2018ApJ...852...72V}, 
while the birth rate of CSO-2s is only 
$\sim 3 \times 10^{-5}\ \mathrm{Gpc^{-3}\ yr^{-1}}$. 
Therefore, even if only a tiny fraction of TDEs trigger jet ignition and produce CSO-2s, the mechanism remains statistically viable, providing further support for the proposed model.

\begin{figure*}
    \centering
    \includegraphics[scale=0.8]{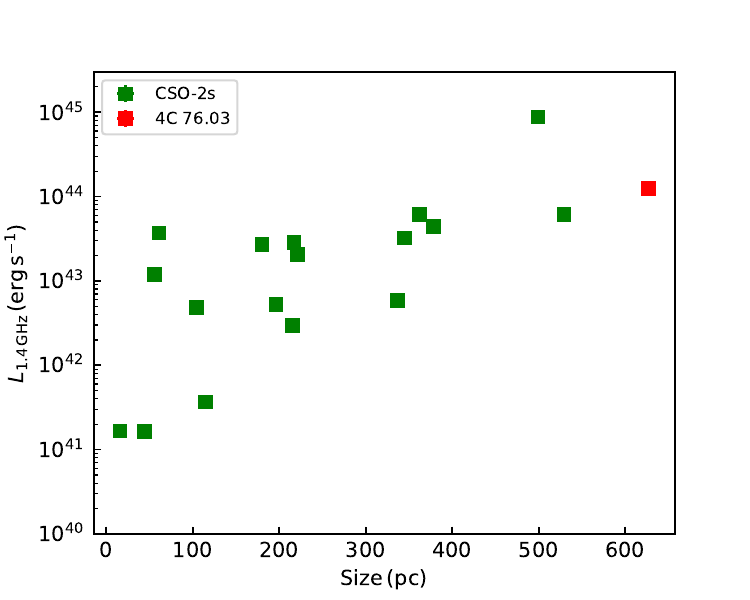}
    \caption{Comparison of the projected linear sizes and 1.4~GHz luminosities. Green points represent 17 CSO-2s, while the red point denotes the CSO2-like source 4C~76.03. The 1.4 GHz data are adopted from \citep{1992ApJS...79..331W}, and the sizes are taken from \citep{2024ApJ...961..241K}.
}
    \label{LS_vs_1.4GHzradio}
\end{figure*}

Figure~\ref{LS_vs_1.4GHzradio} shows the projected linear sizes and 1.4~GHz radio luminosities of the 17 CSO-2s, with 4C~76.03 included for reference. 4C~76.03 exhibits a linear size significantly above the typical cutoff and one of the highest radio luminosities, placing it at the extreme end of the size--luminosity relation. This suggests that efficient or sustained energy injection allows its jet to surpass the usual CSO-2 size limit, potentially enabling evolution into a MSO or even LSO radio galaxy \citep{2024ApJ...961..241K,2024ApJ...977..195D}.

In addition, 4C~76.03 has displayed two pronounced $\gamma$-ray flares lasting $\sim 20$~days, as well as a possible minor flare accompanied by infrared brightening. The short timescales imply that the $\gamma$-ray emission originates from the nuclear core rather than the large-scale radio structure, indicating ongoing significant energy injection that sustains jet activity and the bright arcsecond-scale radio morphology. Consequently, 4C~76.03 appears to be in a transitional phase from a compact radio source toward a classical FR~II radio galaxy.

\begin{figure*}
    \centering
    \includegraphics[scale=0.45]{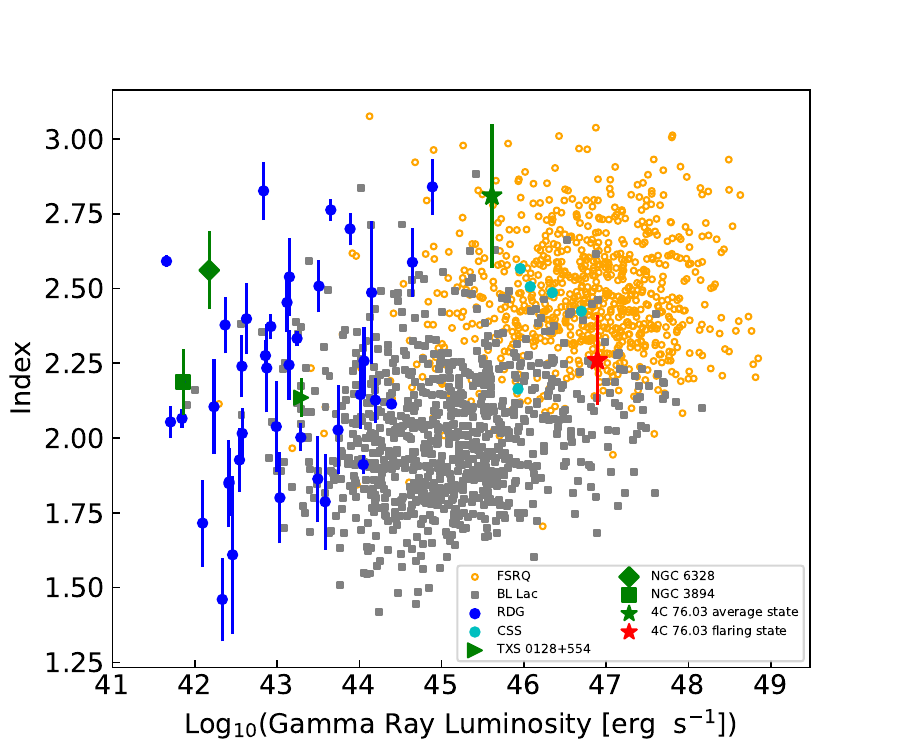}
    \includegraphics[scale=0.45]{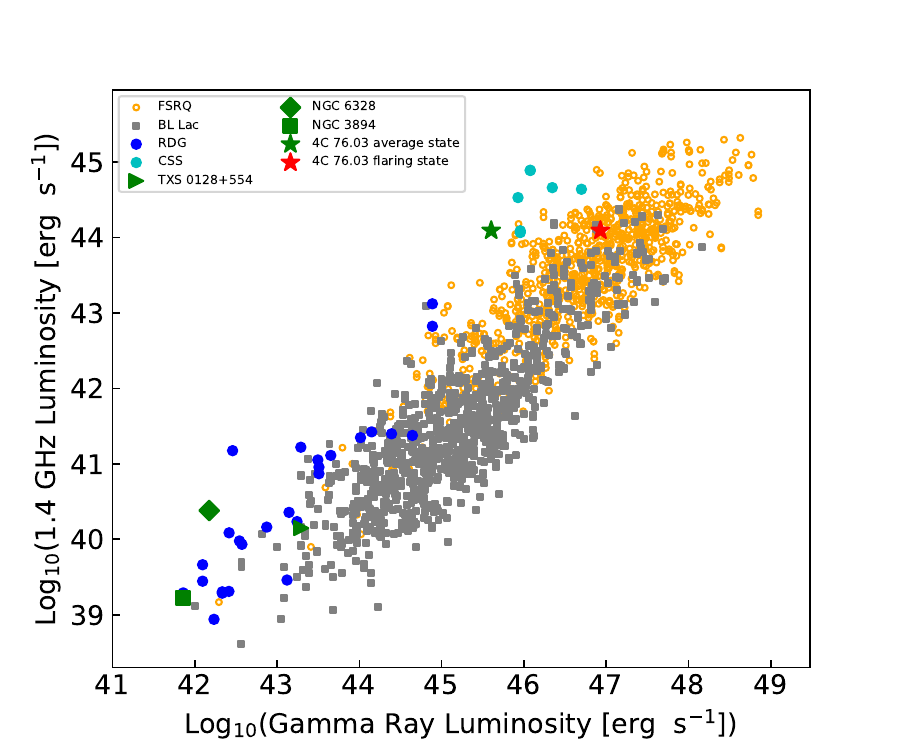}
    \caption{
{\bf Left panel:}
The relationship between the $\gamma$-ray luminosity and the $\gamma$-ray photon index for jetted AGNs with known redshifts included in the 4LAC-DR3 catalog \citep{2022ApJS..263...24A}.
Different classes of jetted AGNs are represented by different symbols: FSRQs (orange points), BL Lacs (gray points), radio galaxies (blue points), and CSS sources (cyan points).
The long-term averaged state (green star) and the short-term flaring state (red star) of 4C~76.03 are also shown.
All $\gamma$-ray luminosities in the figure are K-corrected.
{\bf Right panel:}
The $\gamma$-ray luminosity versus radio luminosity diagram.
The 1.4~GHz radio data are taken from the NRAO VLA Sky Survey (NVSS; \citet{1998AJ....115.1693C}) and the Faint Images of the Radio Sky at Twenty Centimeters (FIRST; \citet{2015ApJ...801...26H}). We did not include DA~362 in our figures, as the origin of its redshift remains uncertain, which could introduce ambiguity in the interpretation \citep{2025ApJ...979...97S}. We also do not show JVAS~J1311+1658, as its redshift is currently unknown \citep{2026ApJ...999..182J}. Similarly, 4C~+39.23B is not shown because its $\gamma$-ray emission is likely contaminated by a flaring source located $\sim$0.1$^\circ$ away \citep{2022ApJ...927..221G}.
}

    \label{lum_vs_index}
\end{figure*}

To place the $\gamma$-ray properties of 4C~76.03 in a broader context, we examine its location relative to the population of extragalactic $\gamma$-ray emitters. Figure~\ref{lum_vs_index} shows the $\gamma$-ray luminosity versus photon spectral index for jetted AGNs with measured redshifts in the 4LAC-DR3 catalog. The distribution is dominated by flat-spectrum radio quasars (FSRQs) and BL Lac objects (BLL), which populate the high-luminosity regime due to strong Doppler boosting in relativistic jets oriented close to the line of sight \citep{2017A&ARv..25....2P,2019ARA&A..57..467B}. 

Classical radio galaxies and confirmed CSOs such as TXS~0128+554, NGC~6328, and NGC~3894 occupy the low-$L_\gamma$ regime, indicating that Doppler boosting plays only a minor role in shaping their $\gamma$-ray emission. In contrast, 4C~76.03 exhibits a photon index comparable to TXS~0128+554 and NGC~3894, but its $\gamma$-ray luminosity is nearly five orders of magnitude higher. This striking discrepancy suggests that the $\gamma$-ray emission in 4C~76.03 arises from a fundamentally different physical origin, likely a newly emerged nuclear jet component oriented close to the line of sight and subject to strong Doppler boosting, distinct from the large-scale, unbeamed radio structure.

The distinction becomes even clearer in the $\gamma$-ray versus radio luminosity plane, where 4C~76.03 occupies an extreme position with both radio and $\gamma$-ray luminosities far exceeding those of young CSOs and even classical radio galaxies. Interestingly, it lies close to five known CSS sources, suggesting that 4C~76.03 may continue its growth along a CSS evolutionary track, gradually developing into a typical CSS source, which is generally considered a young stage of classical FR~II radio galaxies. Unlike most CSO-2s, this indicates that 4C~76.03 may follow a distinct evolutionary pathway, representing a transitional phase from a compact, early-stage jet system to a classical double-lobed radio morphology. As such, 4C~76.03 provides a valuable case for probing the evolutionary connection between CSO-2s, CSS sources, and classical FR~II galaxies, as well as for investigating the physical origin of high-energy emission in young radio jets.

We refer to the behavior observed in some radio galaxies, such as 3C~111 and 3C~120, where superluminal knots ejected from the core are frequently accompanied by intense $\gamma$-ray flares, commonly interpreted as arising from newly ejected jet components oriented closer to the line of sight and thus strongly Doppler boosted \citep{2012ApJ...751L...3G, 2015ApJ...808..162C, 2015ApJ...799L..18T}. An even more striking case is the giant radio galaxy PBC~J2333.9$-$2343, whose core jet appears to have undergone a dramatic reorientation, resulting in a much smaller viewing angle and significantly enhanced high-energy emission \citep{2023MNRAS.525.2187H}. Motivated by these examples and the elevated $\gamma$-ray activity observed in 4C~76.03, we adopt the two-zone leptonic model of \citet{2020ApJ...899....2Z} to interpret its broadband radiation properties.

The emitting blob is assumed to propagate along the jet with a bulk Lorentz factor
$\Gamma$ and to be observed at a viewing angle $\theta$, resulting in a Doppler
beaming factor $\delta = [\Gamma(1-\beta\cos\theta)]^{-1}$. The blob is permeated by
an isotropic magnetic field $B$ and filled with relativistic electrons, which
produce synchrotron radiation as well as synchrotron self-Compton (SSC) and
external Compton (EC) emission. The energy distribution of the electrons is
described by a broken power-law with an exponential cutoff, characterized by a
normalization $K$, minimum, break, and maximum Lorentz factors $\gamma_{\min}$,
$\gamma_{\rm break}$, and $\gamma_{\max}$, and spectral indices $p_1$ and $p_2$
below and above the break, respectively.

For the core region, the radius of the emission zone is constrained by the
variability timescale as $R \leq \delta c t_{\rm obs}/(1+z)$. Throughout this work,
we assume $\delta = \Gamma = 10$ and adopt a representative emission region size
of $R \simeq 5 \times 10^{16}$~cm. Assuming a conical jet geometry with an opening
angle of $\sim 1/\delta$, the distance of the emission region from the central
black hole is estimated as $d \simeq R\delta \approx 0.17$~pc.
We further assume a conservative black hole mass of 
$10^{8}\,M_{\odot}$ and an accretion disk luminosity of 
$L_{\rm disk} \sim 5 \times 10^{44}$~erg~s$^{-1}$.  
If the broad-line region (BLR) reprocesses $\sim10\%$ of the disk luminosity 
($L_{\rm BLR}=0.1\,L_{\rm disk}$), its characteristic radius can be estimated as 
$r_{\rm BLR} = 0.1 (L_{\rm BLR}/10^{45}\ \mathrm{erg\ s^{-1}})^{1/2}$~pc,
yielding $r_{\rm BLR} \approx 0.07$~pc. 
Similarly, assuming that the dust torus reprocesses 
$f_{\rm DT}=0.3$ of the disk luminosity, its characteristic radius is 
$r_{\rm DT} = 2.5 (f_{\rm DT} L_{\rm disk}/10^{45}\ 
\mathrm{erg\ s^{-1}})^{1/2}$~pc,
which gives $r_{\rm DT} \approx 0.97$~pc.  
Since the emission region is located well beyond the BLR but inside the dust
torus, we consider EC scattering of infrared photons from the dust
torus as the dominant source of seed photons for the observed $\gamma$-ray
emission. The dust torus temperature is assumed to be $T_{\rm DT}=1000$~K
\citep{2020Galax...8...72C}. For comparison, the large-scale radio-emitting region is assumed to be only
mildly relativistic, and we therefore adopt $\Gamma = \delta = 1.3$ for this
component.

\begin{figure*}
    \centering
    \includegraphics[scale=0.7]{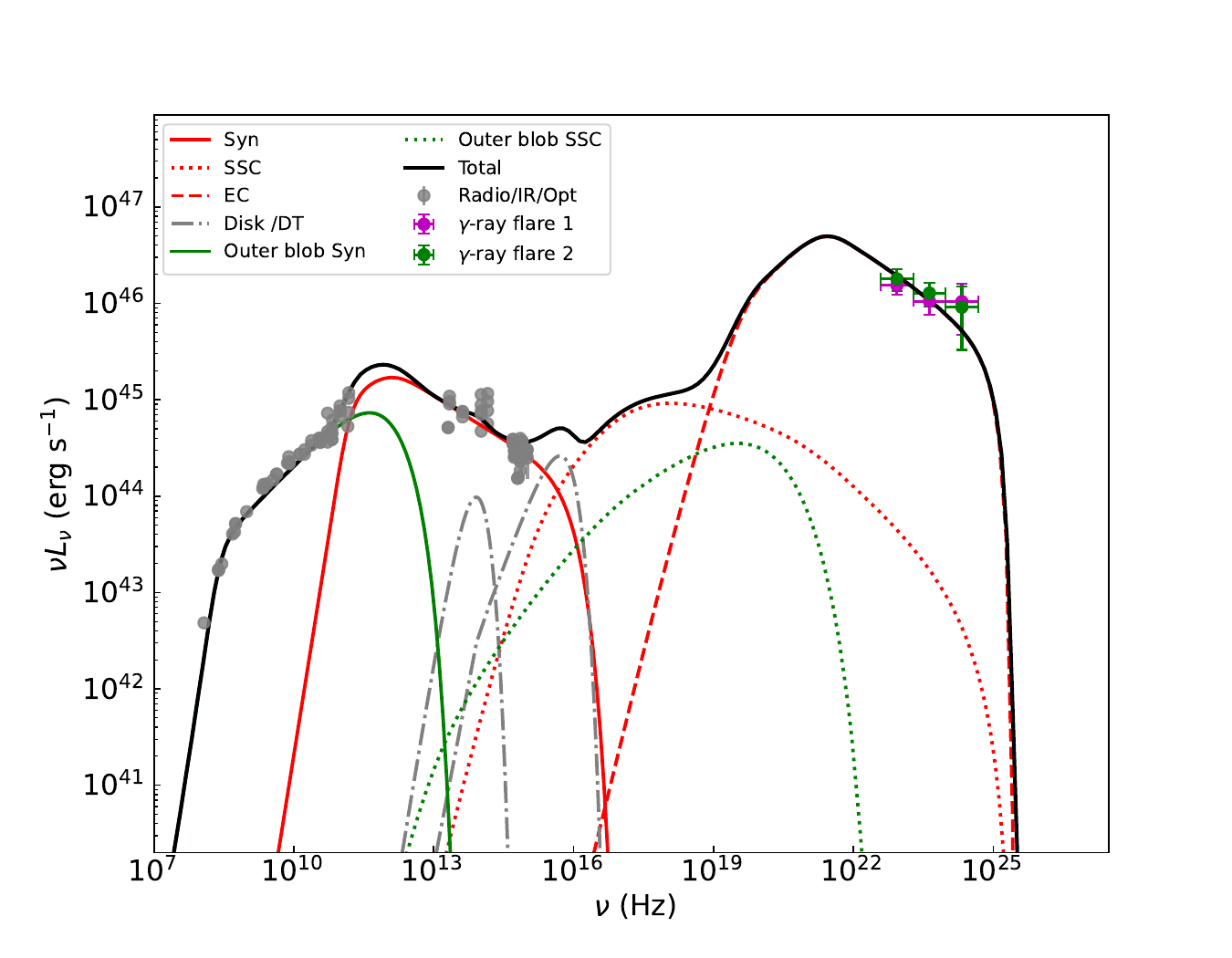}
    \caption{The multi-band SED fitted with a two-zone leptonic model. 
    The red component represents the emission from the core/inner jet, while the green component denotes the emission from the large-scale extended structure. 
    The gray component corresponds to the emission from the dusty torus and the accretion disk. 
    The solid lines indicate synchrotron radiation, the dotted lines represent SSC emission, and the dashed lines denote EC emission.
}
    \label{multband_SED}
\end{figure*}

The resulting SED is shown in Figure~\ref{multband_SED}. The radio emission is predominantly produced by the extended large-scale structures, whereas the infrared–optical and $\gamma$-ray emissions are dominated by the inner core region. Previous modeling of the two other $\gamma$-ray–detected CSOs, NGC~6328 and NGC~3894, suggests that their infrared and optical emission is mainly thermal in origin, arising from the dust torus and the host galaxy stellar population \citep{2022ApJ...941...52S, 2024A&A...684A..65B}. In contrast, the infrared emission of 4C~76.03 is dominated by synchrotron radiation, which may imply ongoing particle acceleration in the core and continued energy injection.

The corresponding model parameters are summarized in Table~\ref{t1}. 
The large ratio $L_{e,k}/L_{B,k} \simeq 37$ indicates that the core-region jet is strongly particle-dominated, 
with relativistic electrons carrying most of the energy budget and efficiently powering the high-energy emission.
However, owing to the lack of simultaneous multiwavelength observations during the $\gamma$-ray flaring episode, 
the physical parameters of the emission region remain weakly constrained. 
The values reported here therefore represent a plausible set obtained through manual exploration of parameter space 
that provides an acceptable fit to the observed SED. 
Future $\gamma$-ray flares accompanied by contemporaneous multiwavelength observations will be essential 
for placing tighter constraints on the emission region and its radiation mechanisms.

In summary, 4C~76.03 displays pronounced $\gamma$-ray variability on
$\sim$20--30 day timescales, indicating that the high-energy emission
originates from a compact region in the nuclear jet and is likely subject
to strong Doppler boosting, and that the core remains actively powered,
requiring sustained energy injection into the jet.
Its extreme $\gamma$-ray and radio luminosities, together with a projected
linear size exceeding the characteristic cutoff of confirmed CSO-2s,
set it apart from typical young radio sources.
These properties indicate that 4C~76.03 may represent a rare transitional
object evolving from a compact CSO-like system toward a CSS source or a
classical FR~II radio galaxy.
Continued monitoring, particularly coordinated multiwavelength observations
during $\gamma$-ray flaring episodes, will be essential for better constraining
the physical conditions of the emission region and for clarifying the
evolution of young radio jets and their high-energy radiation mechanisms.

\section*{Acknowledgments}
We acknowledge the use of several publicly available Python packages, including Astropy \citep{2013A&A...558A..33A}, Matplotlib \citep{hunter2007matplotlib}, Pandas \citep{reback2020pandas}, NumPy, and SciPy \citep{2020NatMe..17..261V}, which were essential for the data analysis and visualization presented in this work. This research has made use of data and software provided by the High Energy Astrophysics Science Archive Research Center (HEASARC), operated by NASA’s Goddard Space Flight Center. We also made use of the NASA/IPAC Extragalactic Database (NED), which is operated by the Jet Propulsion Laboratory, California Institute of Technology, under contract with NASA.

This work was  supported by the National Natural Science Foundation of China (NSFC) under grants No. 12473049.

\bibliography{refs}
\bibliographystyle{aasjournal}


\begin{table}[ht]
    \centering
    \caption{Parameter values used for the SED models}
    \label{t1}
    \begin{tabular}{lcc}
        \toprule
        Parameter & Core/inner jet & Outer blob \\
        \midrule
        $\delta$    & 10   & 1.3   \\
        $R$ (cm)    & $8.6\times10^{16}$   & $5\times10^{20}$   \\
        $B$ (G)    & $1.6\times10^{-1}$   & $1.0\times10^{-3}$   \\
        $\gamma_{\min}$    & 50   & 100 \\
        $\gamma_{\rm break}$    & $4\times10^{2}$   & $1\times10^{4}$ \\
        $\gamma_{\max}$       & $3\times10^{4}$   & $2\times10^{4}$ \\
        $p_{1}$    & 2.0   & 2.0   \\
        $p_{2}$    & 3.7   & 4.0   \\
        $K$ (cm$^{-3}$) & $1.64\times10^{4}$   & $1.42\times10^{-1}$   \\
        \midrule
        $L_{e,k}$ (erg s$^{-1}$)    & $2.5\times10^{45}$   & $2.3\times10^{46}$   \\
        $L_{B,k}$  (erg s$^{-1}$)  & $6.7\times10^{43}$   & $1.6\times10^{45}$   \\
        \bottomrule
    \end{tabular}

    \tablecomments{
    The kinetic luminosity carried by relativistic electrons is defined as
    $L_{e,k} = \pi R^{2}\Gamma^{2} c \, m_{e} c^{2}
    \int N_{e}(\gamma_{e}) \gamma_{e} \, d\gamma_{e}$,
    while the magnetic luminosity is
    $L_{B,k} = \pi R^{2}\Gamma^{2} c \, B^{2}/(8\pi)$.
    }
\end{table}



\end{document}